\documentstyle[preprint,aps,epsf]{revtex}
\begin{document}

\draft

\title{A theory of $\pi /2$ superconducting Josephson junctions}

\author{A.Zyuzin}

\address{Max-Planck-Institut f$\ddot u$r Physik Komplexer Systeme,
01187 Dresden Germany }
\address
{A.F.Ioffe Physical- Technical Institute, 194021, St.Petersburg,
Russia}

\author{B.Spivak}
\address{Physics Department University of Washington, Seattle, WA 98195, USA}

\maketitle

\begin{abstract}
We consider theoretically a Josephson junction with a superconducting critical
current density
which has a random sign along the junction's surface.
We show that the ground state of the junction corresponds to 
the phase difference equal to $\frac{\pi}{2}$. Such a situation can 
take
place in superconductor- ferromagnet-superconductor junctions. 
\end{abstract}


\newpage

The superfluid density connects
supercurrent with superfluid velocity in superconductors. In the framework of BCS theory
 it is a positive quantity. Thus the ground state of a
superconductor corresponds to zero superfluid velocity (See for example
$^{[1]}$).
The question whether and under what circumstances the superfluid density
can
be
negative has been discussed in many theoretical papers. 
a). It has
been shown
$^{[2-5]}$ 
that  if the distribution function of
quasiparticles in superconductors is non-equilibrium the superfluid
density can be negative.
Negative Josephson coupling has been observed experimentally in nonequilibrium
superconductor-normal metal-superconductor junction $^{[6]}$.
b). In the presence of magnetic impurities or resonance states with strong
electron-electron correlation inside the insulator
a superconductor-insulator-superconductor junction could have 
negative Josephson coupling $^{[7-10]}$. 
c). In case of disordered
superconductor-ferromagnet-superconductor junctions the 
 Josephson coupling averaged over
samples has
oscillating sign 
and decays exponentially
with the ferromagnetic layer thickness $L$ $^{[12]}$. Mesoscopic fluctuations of the
coupling, however, do not decay exponentially with $L$. Therefore at large $L$ the
junctions have couplings with random sign $^{[13]}$.  
d). At last, a d-superconductor-normal metal-d-superconductor junction can have
arbitrary sign depending on the relative orientation of the (superconducting)
order
parameters in the superconductors $^{[14,15]}$. The fact that the superfluid
density is negative indicates that the state with zero phase gradient is
unstable.  
The ground state of a Josephson junction with negative critical current
corresponds to a phase difference on the junction which equals $\pi$.

In real situation, however, the sign of the superfluid density can
fluctuate
from point to point inside the sample.
In this paper we consider a Josephson junction, with a coupling
 $J(\bbox{\rho})$ fluctuating from point to
point along the junction's surface and with a random sign.
Here $\bbox{\rho}$ is a
coordinate along the junction's surface.
We show that in this case the ground
state of the system corresponds to a phase difference of the
order parameter of $\frac{\pi}{2}$.

The total energy of the system can be written as 
\begin{equation} E= E_J + \frac {mn_s}{2}
\int d^3\bbox{r}
\biggl(\bbox{v}_{s} (\bbox{r}) \biggr)^2
 \end{equation}  
Here  $m$ is the electron
mass, $\bbox{v}_{s} (\bbox{r}) = \frac{1}{m}\bbox{\nabla}\phi (\bbox{r} )
$ and $\phi (\bbox{r})$ are the superconducting velocity and the phase of
the superconducting order parameter in the bulk of the superconductors,
($\hbar \equiv 1$), $n_s$ is the density of
superconducting electrons, which is assumed to be the same in both
superconductors, $E_J$ is the Josephson energy of the
junction which can be written as 
\begin{equation}
E_J=-\int d^2 \bbox{\rho}  J(\bbox{\rho})
\cos (\chi (\bbox{\rho})), 
\end{equation}
 $\chi(\bbox{\rho})=\phi(z=+0,\bbox{\rho})-\phi(z=-0,\bbox{\rho})$ is
the jump of the phase of
superconducting order
parameter at the junction's surface $z=0$. Here $z$ is the coordinate
perpendicular to the
junction's surface. In Eqs.1,2 we neglected the thickness of the
junction. The second term in Eq.1 corresponds to the energy
associated with supercurrents in the bulk of the superconductors.
We neglect the magnetic field energy in the bulk of the superconductors
assuming that $J(\bbox{\rho})$ is small and the Josephson penetration
length
of the magnetic field along the junction's boundary is large. 
 
In the case, when $J(\bbox{\rho})$ has a random sign, the ground state
of the system corresponds to nonuniform distributions of $\bbox{v}_s
(\bbox{r}
)\neq 0$ and $\chi(\bbox{\rho})\neq 0$. 
The current conservation law in the bulk of
superconductors and at the junction's boundary gives
\begin{equation} 
\Delta \phi(\bbox{r}) = 0
\end {equation}
\begin{equation}  \frac {d\phi (\bbox{\rho},z=+0)}{dz}=
 \frac {d\phi(\bbox{\rho},z=-0)}{dz}=
\frac {m}{n_s} 
J (\bbox{\rho}) \sin (\chi (\bbox{\rho}) ) 
\end {equation}
respectively.
Solutions of Eqs.3,4 determine the extrema of the total energy Eqs.1 and 2.
At given
$\langle\chi(\bbox{\rho})\rangle$ the minimum of Eq.1 corresponds to some
distribution of $\delta \chi(\bbox{\rho})$, which is a random sample
specific quantity. If
$J(\bbox{\rho})$ is small enough, then $\delta
\chi(\bbox{\rho})=\chi(\bbox{\rho})-\langle\chi(\bbox{\rho})\rangle \ll
1$. 
Here the bracket $\langle \rangle$
stands for
averaging over the junction's surface. 

 Substituting $\chi(\bbox{\rho})\sim
\langle\chi(\bbox{\rho})\rangle $ in left hand side of Eq.4 and solving
Eqs.3-4 we obtain an
expression for the minimum of the total junction's energy per unite area, 
\begin{equation}
E_{eff} = -\langle J(\bbox{\rho})\rangle \cos\langle\chi\rangle - J_{eff}\sin^2
\langle\chi\rangle
\end{equation}
as a function of $\langle\chi(\bbox{\rho})\rangle$, where
\begin{equation} 
J_{eff}=\frac {m}{S n_s} \int d\bbox{\rho}_1
d\bbox{\rho}_2 
\frac {\langle \delta I(\bbox{\rho}_1 ) \delta I(\bbox{\rho}_2\rangle}
{2\pi |\bbox{\rho}_1 - \bbox{\rho}_2 |}
\end {equation}

Here $S$ is the area of the junction. 
For $0<2J_{eff}<\langle J\rangle$ the
junction's energy $E_{eff}$ has a minimum
for 
$\langle \chi \rangle=0$,
 while at $2J_{eff}>\langle J\rangle$, the
minimum
corresponds to a nonzero  phase difference
\begin{equation}
\langle \chi \rangle^{min} = \arccos (\frac {\langle J
\rangle}{2J_{eff}}),
\end{equation}
which in the case of $J_{eff}\gg \langle J \rangle$ gives
$\langle \chi\rangle^{min}=\frac{\pi}{2}$.
The derivation presented above is similar to the case of a ferromagnet-normal 
metal-ferromagnet junction with a random in sign exchange energy between the
ferromagnets $^{[16]}$

 Below we consider in more details the case of a superconductor-
disordered ferromagnetic metal-superconductor junction shown in Fig.1.
We will describe the system by the Hamiltonian 
\begin{equation}
 H = H_{BCS} + H_T + H_F.
\end{equation}
where $H_{BCS}$ is the BCS Hamiltonian of superconducting leads,
\begin{equation}
 H_T = t \sum_{i=1,2 ; \alpha} \int_{S_i}\prod_{i} d\bbox{r}_{i} \biggl( 
\Psi^{+}_{i}(\bbox{r}_{i}, \alpha) \Psi_{F} (\bbox{r}, \alpha ) + h.c.  
\biggr). 
\end{equation}
is the Hamiltonian describing the tunneling between the two superconductors
labeled by indexes 1,2,
and the ferromagnet labeled by the index $F$; and $\alpha$ is a spin
index. Furthemore,
 \begin{equation}
 H_F = H_0 + \bbox{H}\bbox{\sigma} 
\end{equation}
is the Hamiltonian of the ferromagnet. The integration in Eq.9 is taken
over the superconductor-ferromagnet surfaces.
The effective magnetic field $\bbox{H}$ models the ferromagnetic exchange
spin splitting,
 $\bbox{\sigma}$ are Pauli matrices; $H_0$ is the hamiltonian of free electron 
gas in a random potential $U(\bbox{r})$.
We assume the following correlation properties of the random 
potential : $<U(\bbox{r})>=0$
and $<U(\bbox{r})U(\bbox{r}')>=\frac {\delta (\bbox{r} - \bbox{r}')}{2\pi
\nu_0 \tau}$.
Here  $\tau$ is the electron elastic mean free time, $\nu_0 $ is
electron density of states at the Fermi level.

Let us introduce a nonlocal Josephson coupling density $J(\bbox{\rho}
,\bbox{\rho}')$ as 
\begin{equation}
E_{J}=\int_{S_{i}} d\bbox{\rho}d\bbox{\rho}' J(\bbox{\rho}
,\bbox{\rho}') \cos(\phi(\bbox{\rho})-\phi(\bbox{\rho}'))
\end{equation}
where $\bbox{\rho}$ and $\bbox{\rho}'$ are coordinates on the surfaces between 
the
ferromagnet
and the two superconductors.
 The integration in Eq.11 is taken over the
surfaces $S$ and $S'$ of the
superconductors.

To lowest order in $t$ we have 
\begin{eqnarray}
J(\bbox{\rho} ,\bbox{\rho}') = \nonumber \\
= \frac{T}{2}\sum_{\epsilon_{n},\alpha}
t^4 \int_{S,S'}  d\bbox{r}  d\bbox{r}'
\biggl[ F^{+}_{\alpha,-\alpha}( \epsilon_n; \bbox{\rho} ;\bbox{r})
G_{\alpha,\alpha}(\epsilon_n ;\bbox{r} , \bbox{\rho}') 
F_{\alpha,-\alpha}(\epsilon_n ; \bbox{\rho}' ;\bbox{r}')
G_{-\alpha,-\alpha}(-\epsilon_n ;\bbox{r}' , \bbox{\rho}) 
+ h.c. \biggr]
\end{eqnarray}
Here $ F^{+}_{\alpha,-\alpha}( \epsilon_n; \bbox{r}_1 \bbox{r}'_1)$
and $ F_{\alpha,-\alpha}( \epsilon_n; \bbox{r}_1 \bbox{r}'_1)$ are
anomalous Green functions
of the
superconducting leads. $G_{\alpha,\alpha}(\epsilon_n ;\bbox{r}_1 ,
\bbox{r}_2)$ are
Green
functions of electron in ferromagnets;
$\epsilon_n = (2n+1)\pi T$ is the fermionic Matsubara frequency.
A derivation identical to Eq.6 gives
\begin{eqnarray}
J_{eff}=\frac {m}{S\ n_s}\int d\bbox{\rho}
d\bbox{\rho}_{1}d\bbox{\rho}'
 d\bbox{\rho}'_{1}
\frac{\langle \delta J(\bbox{\rho} ,\bbox{\rho}_{1})\delta J(\bbox{\rho}'
,\bbox{\rho}'_{1})\rangle}{2\pi |\bbox{\rho}-\bbox{\rho}'|} \nonumber \\
\sim \frac {\xi m}{\pi n_s}\biggl(\frac {g}{8\pi\nu_0 D}\biggr)^4
 \biggl(\frac{D}{2\pi^2 L^2}\biggr)^2
\ln(\frac {\xi}{l}) .
\end{eqnarray}
Here $L$ is the ferromagnet's thickness, $\xi = \sqrt{D/\Delta}$ is the
superconducting coherence length and $\Delta$ is the modulus of the
superconducting order parameter, $g$ is the dimensionless conductance
 of superconductor-ferromagnet boundary per unit area, and $D$ is
the electron diffusion constant, which is assumed to be the same in the
superconductors and in the ferromagnet.
Diagrams which contribute to Eq.13
are
shown in Fig.1. We use the conventional diagram technique for the
averaging
over scattering potential configurations $^{[17]}$ and retain the symbol
$\langle \rangle$ to indicate the averaging over configurations of the
scattering potential.

In the case of no electron reflection from the superconductor-ferromagnet boundary 
 we can
estimate $J_{eff}$ by taking into account that the mesoscopic fluctuations of
the critical current of a SNS junction of the area $L^2$ is of order of
$\frac{D}{L^{2}}$ $^{[11]}$.
 Fluctuations of the Josephson coupling 
associated with different areas of the junction's surface
separated by distances larger than $L$ are uncorrelated.
As a result, we estimate
\begin{equation} 
J_{eff}\sim\frac {m}{n_s L}\biggl(\frac {D}{L^2}\biggr)^2 L^{-2}
\end {equation}
It follows from Eq.12 that $J_{eff}$ decays with $L$ only as a power law.
On the other hand, $\langle J(\bbox{\rho})\rangle $ decreases
exponentially
at
$L>\sqrt{D/H}$ $^{[12]}$ and we have $|\langle J(\bbox{\rho})\rangle |\ll
J_{eff}$.

In conclusion we would like to mention that 
the state discussed above of the superconductor-ferromagnet-superconductor
junction is similar to Fulde-Ferrell-Larkin-Ovchinnikov state
of bulk superconductors $^{[18,19]}$, when in the presence of the
spin splitting energy the 
superconducting ground state corresponds to a nonuniform distribution
of the order parameter. After averaging over the configurations of the impurity
potential the order
parameter in such a state is significantly suppressed by static disorder.
Mesoscopic fluctuations of the order parameter in such a state can however 
survive the disorder.  

Authors kindly acknowledge discussions with P. Fulde.

\newpage

\begin{figure}
  \centerline{\epsfxsize=10cm \epsfbox{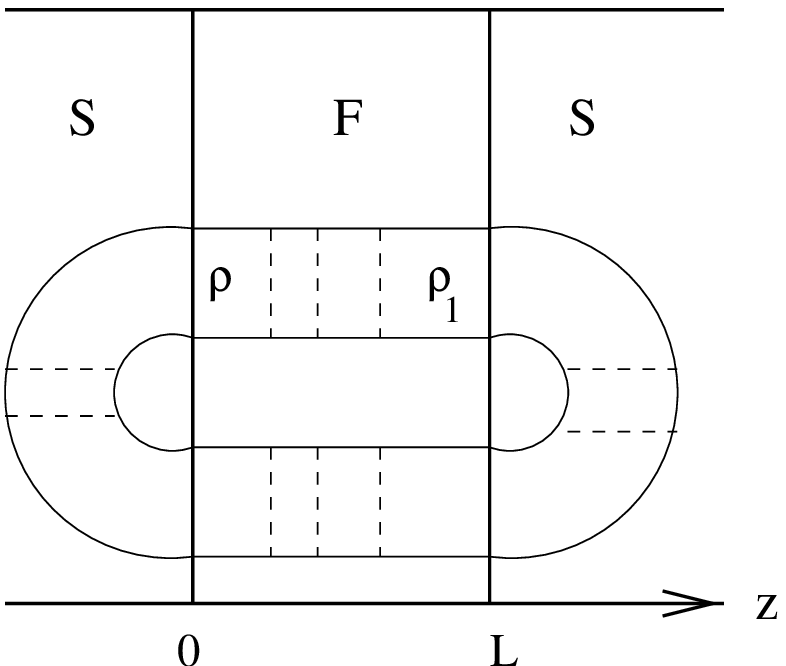}}
  \caption{A schematic picture of the superconductor (S)-ferromagnet (F)-
superconductor (S) junction and
the diagram for calculation of Eq.13. Thin
solid lines inside the ferromagnet correspond to electronic Green
functions $G(\bbox{\rho}, \bbox{\rho}_{1})$, dashed lines correspond to
the elastic scattering, thin solid 
lines inside the superconductors correspond to the anomalous Green
function $F(\bbox{\rho},\bbox{\rho}')$.} \
  \label{fig:fig1}
\end{figure}

\end{document}